\documentclass{revtex4}

\newcommand{\be}{\begin{equation}}
\newcommand{\ee}{\end{equation}}
\newcommand{\ba}{\begin{eqnarray}}
\newcommand{\ea}{\end{eqnarray}}
\newcommand{\ban}{\begin{eqnarray*}}
\newcommand{\ean}{\end{eqnarray*}}

\begin{document}

\title{Plasma Electromagnetic Fluctuations
as an Initial Value Problem}

\author{Stanis\l aw Mr\' owczy\' nski\footnote{Electronic address:
{\tt mrow@fuw.edu.pl}}}

\affiliation{ Institute of Physics, \'Swi\c etokrzyska Academy \\
ul.~\'Swi\c etokrzyska 15, PL - 25-406 Kielce, Poland \\
and So\l tan Institute for Nuclear Studies \\
ul.~Ho\.za 69, PL - 00-681 Warsaw, Poland}

\date{7-th January 2008}

\begin{abstract}

Fluctuations of electric and magnetic fields in the collisionless
plasma are found as a solution of the initial value linearized
problem. The plasma initial state is on average stationary and 
homogeneous. When the state is stable, the initial fluctuations 
decay exponentially and in the long time limit a stationary spectrum 
of fluctuations is established. For the equilibrium plasma it 
reproduces the spectrum obtained from the fluctuation-dissipation 
relation. Fluctuations in the unstable two-stream system are
also discussed.

\end{abstract}

\pacs{52.25.Gj}


\maketitle


\section{Introduction}


Spectrum of electromagnetic fluctuations is an important plasma 
characteristics studied in various contexts. In terrestrial
experiments the spectrum, which is observable through scattering
measurements, signals, for example, an onset of plasma instability
or turbulence. Electromagnetic fluctuations in primordial cosmological 
plasma are analyzed to explain an origin of magnetic fields in the 
Universe.

The fluctuations can be theoretically described using several methods 
reviewed in the classical monographs \cite{Akh75,Sit82}. Modern 
field-theory techniques developed for relativistic plasmas are worked
out in \cite{Siv85,Lemoine:1995fh}. Physically most appealing seems 
to us the method proposed by Rostoker \cite{Ros61} and Klimontovich 
and Silin \cite{Kli62} which is clearly exposed in the handbook 
\cite{LP81}. The method, which is applicable to both equilibrium 
and nonequilibrium plasmas, provides the spectrum of fluctuations as 
a solution of the initial value (linearized) problem. The initial 
plasma state is assumed to be on average stationary and homogeneous. 
When the state is stable, the initial fluctuations are explicitly 
shown to exponentially decay and in the long time limit one finds 
a stationary spectrum of fluctuations. In this way one obtains for 
the equilibrium plasma the spectrum which is alternatively provided 
by the fluctuation-dissipation relation. When the initial state 
is unstable, the memory of initial fluctuations is not lost, as the 
unstable modes, which are present in the initial fluctuation spectrum, 
exponentially grow. 

The fluctuations of the distribution function, electric charge or
longitudinal electric field can be found rather easily, see 
\cite{Ros61,Kli62,LP81}. Analytic computation of the magnetic field
fluctuations appear to be quite lengthy and tedious while the computation 
of fluctuation spectrum of the electric field, which is not constrained 
to be purely longitudinal, is a real challenge even in the collionless 
plasma, as one has to take into account and sort out numerous terms. 
Up to our knowledge such calculations have not been published. In this 
article we study the fluctuation spectrum of electric and magnetic 
fields in detail. In the case of equilibrium, we reproduce the spectrum 
usually provided by the fluctuation-dissipation relation. Fluctuations
in unstable systems are also discussed and, as an example, we compute
the fluctuation spectrum of longitudinal field in the two-stream
system.

The method under consideration, although physically appealing, is 
certainly not the most effective to analyse equilibrium plasmas. And
our actual goal is to set a stage for nonequilibrium calculations
similar to those of the two-stream system. Our particular interests 
is focused on the quark-gluon plasma - a highly relativistic system 
governed by nonAbelian dynamics which, in spite of important differences, 
manifests profound similarities to electromagnetic plasmas discussed at 
length in \cite{Mrowczynski:2007hb}. The quark-gluon plasma produced 
in relativistic heavy-ion collisions is presumably unstable to 
chromomagnetic modes, see the review \cite{Mrowczynski:2005ki}. The 
instability growth is associated with generation of chromomagnetic 
fields which in turn strongly influence transport properties of the 
plasma \cite{Asakawa:2006jn}. The fluctuation spectrum of chromomagnetic 
fields is an important issue to be settled. 

Our paper is organized as follows. In Sec.~\ref{sec-preliminaries} we 
present the theoretical framework to be used in our further considerations. 
The linearized kinetic equation are solved together with Maxwell equations 
by means of the one-sided Fourier transformation in 
Sec.~\ref{sec-initial-value}. The electric and magnetic fields are expressed 
through the initial values of the fields and electron distribution function. 
Sec.~\ref{sec-initial-fluc} deals with the initial fluctuations. 
Those of the distribution function are identified with the fluctuations 
in a classical system of noninteracting particles. The initial fluctuations 
of fields are expressed through the particle fluctuations using the Maxwell 
equations. The well-known fluctuation spectrum of longitudinal electric 
field is  obtained in Sec.~ \ref{sec-fluc-EL} while the fluctuation spectra 
of magnetic and electric fields are derived in Secs.~\ref{sec-fluc-B} and  
\ref{sec-fluc-E}, respectively. It then becomes clear why the analysis of 
longitudinal electric field is much easier than that of the general case. 
In Sec.~\ref{sec-2-streams} we extend our calculations to nonequilibrium 
anisotropic plasma, discussing fluctuations of longitudinal electric 
field in the unstable two-stream system. Our results are summarized and
concluded in Sec.~\ref{sec-discussion}. Throughout the article we use the 
CGS natural units with $c=k_B = 1$.


\section{Preliminaries}
\label{sec-preliminaries}


We consider a classical plasma where ions are assumed to be a passive 
background which merely compensate the charge of electrons. However, 
the ions can be easily included in the considerations. The time 
scale of fluctuations of interest is much shorter than that of 
inter-particle collisions, and consequently the starting point of 
our analysis is the collisionless transport equation of electrons
\be 
\label{transport-eq1}
\Big(\big({\partial \over \partial t} 
+ {\bf v} \cdot \nabla \big) 
- e \big({\bf E}(t, {\bf r}) 
+ {\bf v} \times {\bf B}(t,{\bf r}))\cdot \nabla_p 
\big) \Big) 
f(t, {\bf r},{\bf p}) = 0 \;,
\ee
where $ f(t, {\bf r},{\bf p})$ is the distribution function;
${\bf E}(t, {\bf r})$ and ${\bf B}(t,{\bf r})$ denote 
the electric and magnetic fields in the plasma. 

The transport equation (\ref{transport-eq1}) is supplemented
by the Maxwell equations
\ba
\label{Maxwell-eqs-x}
\nabla \cdot {\bf E}(t, {\bf r}) &=& 4\pi \rho (t, {\bf r}) 
\;,\;\;\;\;\;\;\;\;\;\;\;\;\;
\nabla \cdot {\bf B}(t, {\bf r}) = 0 \;, \\[2mm]
\nonumber
\nabla \times {\bf E}(t, {\bf r}) &=& - 
{\partial {\bf B}(t, {\bf r}) \over \partial t} \;,
\;\;\;\;\;
\nabla \times {\bf B}(t, {\bf r}) = 
4\pi {\bf j}(t, {\bf r}) +{\partial {\bf E}(t, {\bf r}) \over \partial t} \;,
\ea
with the electric charge density and current given as
\ba
\rho (t, {\bf r}) &=& - e \int {d^3p \over (2\pi)^3}
f(t,{\bf r},{\bf p}) + e \, n_{\rm ions} \;, \\[2mm]
{\bf j}(t, {\bf r}) &=& - e \int {d^3p \over (2\pi)^3} \,
{\bf v} \, f(t,{\bf r},{\bf p}) \;.
\ea

The distribution function is assumed to be of the form
\be
f(t, {\bf r},{\bf p}) = f^0({\bf p}) + \delta f(t, {\bf r},{\bf p})
\;,
\ee
with 
\be
\label{smallness}
f^0({\bf p}) \gg \delta f(t, {\bf r},{\bf p}) 
\;,\;\;\;\;\;\;
|\nabla_p f^0({\bf p})| \gg 
|\nabla_p \delta f(t, {\bf r},{\bf p})| \;,
\ee
and
\be
\int {d^3p \over (2\pi)^3} f^0({\bf p}) - n_{\rm ions} = 0
\;, \;\;\;\;\;\;\;
\int {d^3p \over (2\pi)^3} \, {\bf v} \, f^0(t,{\bf r},{\bf p}) = 0\;.
\ee

The transport equation linearized in $\delta f$ is
\be 
\label{lin-trans-eq1}
\big({\partial \over \partial t} 
+ {\bf v} \cdot \nabla \big) \delta f(t, {\bf r},{\bf p})
- e \big({\bf E}(t, {\bf r}) 
+ {\bf v} \times {\bf B}(t,{\bf r})\big)\cdot 
\nabla_p f^0({\bf p}) = 0 \;.
\ee
and
\ba
\rho (t, {\bf r}) &=& - e \int {d^3p \over (2\pi)^3}
\delta f(t,{\bf r},{\bf p})
\;, \\[2mm]
\label{current-lin}
{\bf j}(t, {\bf r}) &=& - e \int {d^3p \over (2\pi)^3} \,
{\bf v} \, \delta f(t,{\bf r},{\bf p}) \;.
\ea


\section{Initial value problem}
\label{sec-initial-value}


We are going to solve the linearized transport equation 
(\ref{lin-trans-eq1}) and Maxwell equations (\ref{Maxwell-eqs-x}) 
with the initial conditions
\be
\delta f(t=0,{\bf r},{\bf p}) = \delta f_0({\bf r},{\bf p})
\;,\;\;\;\;\;
{\bf E}(t=0,{\bf r}) = {\bf E}_0({\bf r})
\;,\;\;\;\;\;
{\bf B}(t=0,{\bf r})= {\bf B}_0({\bf r}) \;.
\ee
We apply to the equations the one-sided Fourier transformation
defined as
\be
f(\omega,{\bf k}) = \int_0^\infty dt \int d^3r 
e^{i(\omega t - {\bf k}\cdot {\bf r})}
f(t,{\bf r}) \;.
\ee
The inverse transformation is 
\be
f(t,{\bf r}) = \int_{-\infty +i\sigma}^{\infty +i\sigma}
{d\omega \over 2\pi} \int {d^3k \over (2\pi)^3} 
e^{-i(\omega t - {\bf k}\cdot {\bf r})} f(\omega,{\bf k}) \;,
\ee
where the real parameter $\sigma > 0$ is chosen is such a
way that the integral over $\omega$ is taken along a straight
line in the complex $\omega-$plane, parallel to the real
axis, above all singularities of $f(\omega,{\bf k})$.

We note that 
\be
\int_0^\infty dt \int d^3r 
e^{i(\omega t - {\bf k}\cdot {\bf r})}
{\partial f(t,{\bf r}) \over \partial t}
= -i\omega f(\omega,{\bf k}) - f(t=0,{\bf k}) \;.
\ee

The linearized transport (\ref{lin-trans-eq1}) and Maxwell 
equations (\ref{Maxwell-eqs-x}), which are transformed by means 
of the one-sided Fourier transformation, read
\be
-i(\omega - {\bf k}\cdot {\bf v}) \delta f(\omega,{\bf k},{\bf p})
- e \big({\bf E}(\omega,{\bf k}) 
+ {\bf v} \times {\bf B}(\omega,{\bf k})\big)\cdot 
\nabla_p f^0({\bf p}) =   \delta f_0({\bf k},{\bf p})\;,
\ee
\ba
\label{Maxwell-eqs-k}
i {\bf k} \cdot {\bf E}(\omega,{\bf k}) 
= 4\pi \rho (\omega,{\bf k}) 
\;,\;\;\;\;\;\;\;\;\;\;\;\;\;
i {\bf k} \cdot {\bf B}(\omega,{\bf k}) &=& 0 \;, 
\\[2mm] \nonumber
i {\bf k} \times {\bf E}(\omega,{\bf k}) 
= i\omega {\bf B}(\omega,{\bf k}) + {\bf B}_0({\bf k}) \;,
\;\;\;\;\;
i {\bf k} \times {\bf B}(\omega,{\bf k}) 
&=& 
4\pi {\bf j}(\omega,{\bf k}) 
-i\omega {\bf E}(\omega,{\bf k}) - {\bf E}_0({\bf k}) \;.
\ea

One finds the solution of the transport equation as
\be
\label{solution1}
\delta f(\omega,{\bf k},{\bf p})
= \frac{i}{\omega - {\bf k}\cdot {\bf v}}
\Big( e \big({\bf E}(\omega,{\bf k})
+ {\bf v} \times {\bf B}(\omega,{\bf k})\big)\cdot
\nabla_p f^0({\bf p}) +  \delta f_0({\bf k},{\bf p})\Big)\;.
\ee

\subsection{Electric field}

Substituting the solution (\ref{solution1}) into the Fourier
transformed current (\ref{current-lin}) and using the third Maxwell 
equation (\ref{Maxwell-eqs-k}) to express the magnetic field 
through the electric one, the current gets the form
\ba
{\bf j}(\omega,{\bf k}) &=& - ie^2 \int {d^3p \over (2\pi)^3} \,
\frac{{\bf v}}{\omega - {\bf v}\cdot {\bf k}}
\Big(\big(1-\frac{{\bf k}\cdot {\bf v}}{\omega}\big)
{\bf E}(\omega,{\bf k})
+ \frac{1}{\omega}\big({\bf v} \cdot {\bf E}(\omega,{\bf k})\big) {\bf k}
\Big)\cdot \nabla_p f^0({\bf p})
\\[2mm]
&+& 
e^2 \int {d^3p \over (2\pi)^3} \,
\frac{{\bf v}}{\omega - {\bf v}\cdot {\bf k}}
\Big(\frac{1}{\omega} {\bf v}\times {\bf B}_0({\bf k}) \Big)\cdot
\nabla_p f^0({\bf p})
- ie 
\int {d^3p \over (2\pi)^3} \,
\frac{{\bf v}}
{\omega - {\bf k}\cdot {\bf v}}\, \delta f_0({\bf k},{\bf p})
\;.
\ea

Since the dielectric tensor $\varepsilon^{ij}(\omega,{\bf k})$
in the collisionless limit equals \cite{Akh75}
\be
\varepsilon^{ij}(\omega,{\bf k}) = \delta^{ij} +
\frac{4\pi e^2}{\omega} \int {d^3p \over (2\pi)^3} \,
\frac{v^i}{\omega - {\bf v}\cdot {\bf k}+i0^+}
\Big(\big(1-\frac{{\bf k}\cdot {\bf v}}{\omega}\big)
\delta^{jk}
+ \frac{v^jk^k}{\omega} \Big) \nabla_p^k f^0({\bf p})  \;,
\ee
the current can be written as 
\ba
\label{current2}
j^i(\omega,{\bf k}) &=& 
\frac{-i\omega}{4\pi}
\big(\varepsilon^{ij}(\omega,{\bf k}) - \delta^{ij} \big)
E^j(\omega,{\bf k}) 
\\[2mm]\nonumber
&+&
e^2 \int {d^3p \over (2\pi)^3} \,
\frac{v^i}{\omega - {\bf v}\cdot {\bf k}}
(\Big(\frac{1}{\omega} {\bf v}\times {\bf B}_0({\bf k})\Big)^j
\nabla_p^j f^0({\bf p})
- ie \int {d^3p \over (2\pi)^3} \,
\frac{v^i}
{\omega - {\bf k}\cdot {\bf v}}\, \delta f_0({\bf k},{\bf p})
\;.
\ea

Combing the third and fourth Maxwell equations (\ref{Maxwell-eqs-k}), 
one finds
\be
\label{eq3}
\big[(\omega^2 - {\bf k}^2) \, \delta^{ij}
+ k^ik^j \big] E^j(\omega,{\bf k})
= - 4\pi i\omega \, j^i(\omega,{\bf k})
+i \omega E_0^i({\bf k})
-i \big({\bf k} \times {\bf B}_0({\bf k})\big)^i \;.
\ee
Substituting the current (\ref{current2}) into Eq.~(\ref{eq3}),
one obtains
\ba
\label{E-field2}
\big[ - {\bf k}^2 \delta^{ij} + k^ik^j 
+ \omega^2 \varepsilon^{ij}(\omega,{\bf k}) \big] E^j(\omega,{\bf k})
= - 4\pi i e^2 \int {d^3p \over (2\pi)^3} \,
\frac{v^i}{\omega - {\bf v}\cdot {\bf k}}
\big({\bf v}\times {\bf B}_0({\bf k})\big)^j
\nabla_p^j f^0({\bf p})
\\[2mm] \nonumber 
- 4\pi e \omega \int {d^3p \over (2\pi)^3} \,
\frac{v^i}
{\omega - {\bf k}\cdot {\bf v}}\, \delta f_0({\bf k},{\bf p})
+i \omega E_0^i({\bf k})
-i \big({\bf k} \times {\bf B}_0({\bf k})\big)^i \;.
\ea

Denoting the matrix in left-hand-side of Eq.~(\ref{E-field2}) as
\be
\label{matrix-sigma}
\Sigma^{ij}(\omega,{\bf k}) \equiv
- {\bf k}^2 \delta^{ij} + k^ik^j 
+ \omega^2 \varepsilon^{ij}(\omega,{\bf k}) \;,
\ee
the electric field given by Eq.~(\ref{E-field2}) can be written 
down as
\ba
\label{E-field-final}
E^i(\omega,{\bf k})
= &-& 4\pi e \int {d^3p \over (2\pi)^3} \,
\frac{(\Sigma^{-1})^{ij}(\omega,{\bf k})v^j}
{\omega - {\bf v}\cdot {\bf k}} 
\Big[ 
i e\big({\bf v}\times {\bf B}_0({\bf k})\big) \cdot
\nabla_p f^0({\bf p})
+ \omega \delta f_0({\bf k},{\bf p}) \Big]
\\[2mm] \nonumber 
&+& 
 i \omega (\Sigma^{-1})^{ij}(\omega,{\bf k}) E_0^j({\bf k})
-i (\Sigma^{-1})^{ij}(\omega,{\bf k}) 
\big({\bf k} \times {\bf B}_0({\bf k})\big)^j \;,
\ea
which is the main result of this section.

When the plasma stationary state described by $f^0({\bf p})$
is isotropic, the dielectric tensor can be expressed through
its longitudinal and transverse components
\be
\varepsilon^{ij}(\omega,{\bf k}) = 
\varepsilon_L(\omega,{\bf k}) \: \frac{k^ik^j}{{\bf k}^2}
+ \varepsilon_T(\omega,{\bf k}) \:
\Big(\delta^{ij} - \frac{k^ik^j}{{\bf k}^2}\Big) \;,
\ee
where $\varepsilon_L(\omega,{\bf k})$ and  
$\varepsilon_T(\omega,{\bf k})$ are well known \cite{LP81}
to be equal to
\be
\label{eL}
\varepsilon_L(\omega,{\bf k}) = 1+ \frac{4\pi e^2}{{\bf k}^2}
\int {d^3p \over (2\pi)^3} \frac{1}
{\omega - {\bf k} \cdot {\bf v}+i0^+}
{\bf k} \cdot \frac{\partial f^0({\bf p})}{\partial{\bf p}} \;,
\ee
\be
\label{eT}
\varepsilon_T(\omega,{\bf k}) = 1+ \frac{2\pi e^2}{\omega}
\int {d^3p \over (2\pi)^3} \frac{1}
{\omega - {\bf k} \cdot {\bf v}+i0^+}
\bigg[
{\bf v} \cdot \frac{\partial f^0({\bf p})}{\partial{\bf p}} 
- \frac{{\bf k} \cdot {\bf v}}{{\bf k}^2}
{\bf k} \cdot \frac{\partial f^0({\bf p})}{\partial{\bf p}} 
\bigg]\;.
\ee
The matrix $\Sigma^{ij}(\omega,{\bf k})$, which then equals
\be
\Sigma^{ij}(\omega,{\bf k}) = 
\omega^2\varepsilon_L(\omega,{\bf k})
\frac{k^ik^j}{{\bf k}^2}
+ \big(\omega^2 \varepsilon_T(\omega,{\bf k})-{\bf k}^2\big)
\Big(\delta^{ij} - \frac{k^ik^j}{{\bf k}^2}\Big) \;,
\ee
can be inverted as 
\be
\label{inv-sigma}
(\Sigma^{-1})^{ij}(\omega,{\bf k}) = 
\frac{1}{\omega^2 \varepsilon_L(\omega,{\bf k})}
\frac{k^ik^j}{{\bf k}^2}
+ \frac{1}{\omega^2 \varepsilon_T(\omega,{\bf k})-{\bf k}^2}
\Big(\delta^{ij} - \frac{k^ik^j}{{\bf k}^2}\Big) \;.
\ee

When the momentum distribution $f^0({\bf p})$ is isotropic, 
$\nabla_p f^0({\bf p}) \sim {\bf p}$, and consequently
$\big({\bf v}\times {\bf B}_0({\bf k})\big) \cdot
\nabla_p f^0({\bf p}) = 0$. Therefore, the first term in
the right-hand-side of Eq.~(\ref{E-field-final}) vanishes
and the electric field is found as
\ba
\label{E-field-final-iso}
E^i(\omega,{\bf k})
= &-& 4\pi e \omega \Bigg(
\frac{1}{\omega^2 \varepsilon_L(\omega,{\bf k})}
\frac{k^ik^j}{{\bf k}^2}
+ \frac{1}{\omega^2 \varepsilon_T(\omega,{\bf k})-{\bf k}^2}
\Big(\delta^{ij} - \frac{k^ik^j}{{\bf k}^2}\Big)
\Bigg)\int {d^3p \over (2\pi)^3} \,
\frac{v^j}
{\omega - {\bf k}\cdot {\bf v}}\, \delta f_0({\bf k},{\bf p})
\\[2mm]\nonumber 
&+& i \omega \Bigg(
\frac{1}{\omega^2 \varepsilon_L(\omega,{\bf k})}
\frac{k^ik^j}{{\bf k}^2}
+ \frac{1}{\omega^2 \varepsilon_T(\omega,{\bf k})-{\bf k}^2}
\Big(\delta^{ij} - \frac{k^ik^j}{{\bf k}^2}\Big)
\Bigg) E_0^j({\bf k})
- \frac{i\big({\bf k} \times {\bf B}_0({\bf k})\big)^i}
{\omega^2 \varepsilon_T(\omega,{\bf k})-{\bf k}^2} \;.
\ea

If the field is purely longitudinal, 
$$
{\bf E}(\omega,{\bf k}) = 
\big({\bf k}\cdot {\bf E}(\omega,{\bf k})\big) \,
\frac{\bf k}{{\bf k}^2} 
\;,\;\;\;\;
{\bf E}_0({\bf k}) = 
\big({\bf k}\cdot {\bf E}_0({\bf k})\big) \,
\frac{\bf k}{{\bf k}^2} \;,
$$
Eq.~(\ref{E-field-final-iso}) gives
\ba
\label{E-field-final-iso-long}
{\bf k}\cdot {\bf E}(\omega,{\bf k}) 
= - \frac{4\pi e}{\omega \varepsilon_L(\omega,{\bf k})}
\int {d^3p \over (2\pi)^3} \,
\frac{{\bf k}\cdot {\bf v}}
{\omega - {\bf k}\cdot {\bf v}}\, \delta f_0({\bf k},{\bf p})
+ \frac{i{\bf k}\cdot {\bf E}_0({\bf k})}
{\omega \varepsilon_L(\omega,{\bf k})}\;.
\ea

Taking into account that 
$$ 
i{\bf k}\cdot {\bf E}_0({\bf k}) = 4\pi \rho_0({\bf k})
= -4\pi e \int {d^3p \over (2\pi)^3} \,
\delta f_0({\bf k},{\bf p}) \;,
$$
Eq.~(\ref{E-field-final-iso-long}) can be rewritten as
\ba
\label{E-field-final-iso-long2}
{\bf k}\cdot {\bf E}(\omega,{\bf k}) 
= - \frac{4\pi e}{\varepsilon_L(\omega,{\bf k})}
\int {d^3p \over (2\pi)^3} \,
\frac{\delta f_0({\bf k},{\bf p})}
{\omega - {\bf k}\cdot {\bf v}}
\;.
\ea
Eq.~(\ref{E-field-final-iso-long2}) can be obtained directly 
by substituting the solution of transport equation 
(\ref{solution1}) (with ${\bf B}=0$) into the first 
Maxwell equation. Then, the initial electric field
does not show up.

\subsection{Magnetic field}

Using again the third Maxwell equation (\ref{Maxwell-eqs-k}) to 
express the magnetic field through the electric one, 
Eq.~(\ref{E-field-final}) immediately provides
\ba
\label{B-field-final}
B^i(\omega,{\bf k}) = \frac{1}{\omega}
\epsilon^{ijk}k^j
(\Sigma^{-1})^{kl}(\omega,{\bf k})
\Bigg( - 4\pi i e^2 \int {d^3p \over (2\pi)^3} \,
\frac{v^l}{\omega - {\bf v}\cdot {\bf k}}
\big({\bf v}\times {\bf B}_0({\bf k})\big) \cdot
\nabla_p f^0({\bf p})
\\[2mm] \nonumber 
- 4\pi e \omega \int {d^3p \over (2\pi)^3} \,
\frac{v^l}
{\omega - {\bf k}\cdot {\bf v}}\, \delta f_0({\bf k},{\bf p})
+i \omega E_0^l({\bf k})
-i \big({\bf k} \times {\bf B}_0({\bf k})\big)^l  
\Bigg) + \frac{i}{\omega}B_0^i({\bf k})
\;.
\ea
When the plasma stationary state is isotropic and 
$(\Sigma^{-1})^{ij}(\omega,{\bf k})$ is given by Eq.~(\ref{inv-sigma}),
one finds
\be\epsilon^{ijk}k^j
(\Sigma^{-1})^{kl}(\omega,{\bf k}) = 
\frac{\epsilon^{ijl}k^j}
{\omega^2 \varepsilon_T(\omega,{\bf k})-{\bf k}^2}
\;.
\ee
The first term in the right-hand-side of Eq.~(\ref{B-field-final})
vanishes, because $\big({\bf v}\times {\bf B}_0({\bf k})\big) \cdot
\nabla_p f^0({\bf p}) =0$, and thus
\ba
\label{B-field-final-iso}
{\bf B}(\omega,{\bf k}) = 
&-& \frac{4\pi e }
{\omega^2 \varepsilon_T(\omega,{\bf k})-{\bf k}^2}
\int {d^3p \over (2\pi)^3} \,
\frac{{\bf k} \times {\bf v}}{\omega - {\bf k}\cdot {\bf v}}\, 
\delta f_0({\bf k},{\bf p})
+ \frac{i{\bf k} \times {\bf E_0}({\bf k})}
{\omega^2 \varepsilon_T(\omega,{\bf k})-{\bf k}^2}
+ \frac{i\omega \varepsilon_T(\omega,{\bf k})}
{\omega^2 \varepsilon_T(\omega,{\bf k})-{\bf k}^2}
{\bf B}_0({\bf k})
\;.
\ea


\section{Initial Fluctuations}
\label{sec-initial-fluc}


The correlation functions of electric or magnetic fields, 
$\langle E^i(t_1,{\bf r}_1) E^j(t_2,{\bf r}_2) \rangle$,
$\langle B^i(t_1,{\bf r}_1) B^j(t_2,{\bf r}_2) \rangle$
($\langle \cdots \rangle$ denotes averaging over statistical
ensemble), are determined by the fields ${\bf E}(t,{\bf r})$, 
${\bf B}(t,{\bf r})$ found in the previous section and 
the initial correlations 
$\langle \delta f_0({\bf r}_1,{\bf p}_1) 
\delta f_0({\bf r}_2,{\bf p}_2)\rangle$, 
$\langle E_0^i({\bf r}_1) E_0^j({\bf r}_2) \rangle$,
$\langle B_0^i({\bf r}_1) B_0^j({\bf r}_2) \rangle$,
$\langle \delta f_0({\bf r}_1,{\bf p}_1) 
E_0^j({\bf r}_2) \rangle$, 
$\langle \delta f_0({\bf r}_1,{\bf p}_1) 
B_0^j({\bf r}_2) \rangle$, and  
$\langle E_0^i({\bf r}_1) B_0^j({\bf r}_2) \rangle$
which are discussed in this section. 

We identify the initial correlation function 
$\langle \delta f_0({\bf r}_1,{\bf p}_1) 
\delta f_0({\bf r}_1,{\bf p}_1)\rangle$
with the correlation function 
$\langle \delta f(t_1,{\bf r}_1,{\bf p}_1) 
\delta f(t_2,{\bf r}_2,{\bf p}_2)\rangle_{\rm free}$
taken at $t_1 = t_2 = 0$ of the system of free classical 
particles (obeying Boltzmann statistics) in a stationary 
homogeneous state described by the distribution function 
$f^0({\bf p})$. As well known \cite{LP81},
\be
\langle \delta f(t_1,{\bf r}_1,{\bf p}_1) 
\delta f(t_2,{\bf r}_2,{\bf p}_2)\rangle_{\rm free}
= (2\pi )^3 \delta^{(3)}({\bf p}_2 - {\bf p}_1) \,
\delta^{(3)}\big({\bf r}_2 - {\bf r}_1 
- {\bf v}_1(t_2 - t_1)\big) \: f^0({\bf p}_1) \;.
\ee
Then,
\be
\langle \delta f_0({\bf r}_1,{\bf p}_1) 
\delta f_0({\bf r}_1,{\bf p}_1)\rangle
= \langle \delta f(t_1=0,{\bf r}_1,{\bf p}_1) 
\delta f(t_2=0,{\bf r}_2,{\bf p}_2)\rangle_{\rm free}
= (2\pi )^3 \delta^{(3)}({\bf p}_2 - {\bf p}_1) \,
\delta^{(3)}({\bf r}_2 - {\bf r}_1) \: f^0({\bf p}_1) \;,
\ee
and
\be
\label{ff-0}
\langle \delta f_0({\bf k}_1,{\bf p}_1) 
\delta f_0({\bf k}_2,{\bf p}_2) \rangle
= (2\pi )^3\delta^{(3)}({\bf p}_2 - {\bf p}_1) \:
(2\pi )^3 \delta^{(3)}({\bf k}_2 + {\bf k}_1) \: f^0({\bf p}_1) \;.
\ee

The correlation function 
$k_2^j \langle \delta f_0({\bf k}_1,{\bf p}_1) E_0^j({\bf k}_2)\rangle$
can be also expressed through $\langle \delta f_0({\bf k}_1,{\bf p}_1)
\delta f_0({\bf k}_2,{\bf p}_2)\rangle$. Using the first Maxwell equation, 
one finds
\ba
\label{fEL-0}
k_2^j \langle \delta f_0({\bf k}_1,{\bf p}_1) 
E_0^j({\bf k}_2)\rangle 
&=& -4\pi i
\langle \delta f_0({\bf k}_1,{\bf p}_1)
\rho_0 ({\bf k}_2)\rangle 
\\[2mm] \nonumber
&=&  4\pi i e \int {d^3p_2 \over (2\pi)^3} \,
\langle \delta f_0({\bf k}_1,{\bf p}_1)
\delta f_0({\bf k}_2,{\bf p}_2)\rangle 
=  4\pi i e (2\pi )^3
\delta^{(3)}({\bf k}_2 + {\bf k}_1) \: f^0({\bf p}_1).
\ea
And finally,
\ba
\label{ELEL-0}
k_1^ik_2^j \langle E_0^i({\bf k}_1) E_0^j({\bf k}_2) \rangle
&=& -16\pi^2
\langle \rho_0({\bf k}_1) \rho_0 ({\bf k}_2)\rangle
=-16\pi^2 e^2 \int {d^3p_1 \over (2\pi)^3} \,
{d^3p_2 \over (2\pi)^3} \,
\langle \delta f_0({\bf k}_1,{\bf p}_1)
\delta f_0({\bf k}_2,{\bf p}_2)\rangle
\\[2mm] \nonumber
&=& -16\pi^2 e^2 
(2\pi)^3\delta^{(3)}({\bf k}_2 + {\bf k}_1) 
\int {d^3p \over (2\pi)^3} \, f^0({\bf p}) \;.
\ea

When the electric field is not purely longitudinal, the 
computation of the initial correlations 
$\langle E_0^i({\bf r}_1) E_0^j({\bf r}_2) \rangle$,
$\langle \delta f_0({\bf r}_1,{\bf p}_1) 
E_0^j({\bf r}_2) \rangle$ is more complicated, as 
the electric field ${\bf E}_0({\bf r})$ is not fully
determined by $\delta f_0({\bf r},{\bf p})$ but 
$\delta f(t,{\bf r},{\bf p})$ enters here.
To compute 
$\langle E_0^i({\bf r}_1) E_0^j({\bf r}_2) \rangle$,
$\langle \delta f_0({\bf r}_1,{\bf p}_1) 
E_0^j({\bf r}_2) \rangle$ as well as 
$\langle B_0^i({\bf r}_1) B_0^j({\bf r}_2) \rangle$,
$\langle \delta f_0({\bf r}_1,{\bf p}_1) 
B_0^j({\bf r}_2) \rangle$, and  
$\langle E_0^i({\bf r}_1) B_0^j({\bf r}_2) \rangle$,
we use the Maxwell equations transformed using the
Fourier transformation not the one-sided Fourier 
transformation. Actually, the Fourier transformed Maxwell
equations are very similar to the one-sided Fourier 
transformed Maxwell equations (\ref{Maxwell-eqs-k}).
The initial electric and magnetic fields are simply
absent in the former ones. However, it should be clearly 
stated that the one-sided Fourier transformation is 
{\em not} mixed up with the Fourier transformation. The 
latter is used to compute only the initial fluctuations 
which are independent of $\omega$.

Combing the third and the fourth Maxwell equation, one 
gets the equation as Eq.~(\ref{E-field2}) but the terms
with ${\bf E}_0({\bf k})$ and ${\bf B}_0({\bf k})$ are
absent. Inverting the matrix in the right-hand-side of 
the equation, we get the electric field expressed through
the current
\be
E^i(\omega,{\bf k})
= - 4\pi i\omega 
\bigg[ \frac{1}{\omega^2} 
\frac{k^ik^j}{{\bf k}^2} + 
\frac{1}{\omega^2 - {\bf k}^2} \, \Big(
\delta^{ij} - \frac{k^ik^j}{{\bf k}^2}\Big)
\bigg] \, j^j(\omega,{\bf k}) \;.
\ee
The magnetic field is given as
\be
\label{B-field-2}
{\bf B}(\omega,{\bf k})
= - \frac{4\pi i}{\omega^2 - {\bf k}^2} \;  
{\bf k} \times {\bf j}(\omega,{\bf k})\;.
\ee

The correlation function 
$\langle E_0^i({\bf k}_1) E_0^j({\bf k}_2) \rangle$
is derived as  
\ba
\langle E_0^i({\bf k}_1) E_0^j({\bf k}_2) \rangle
&=& \int \frac{d\omega_1}{2\pi}\frac{d\omega_2}{2\pi}
\langle E^i(\omega_1,{\bf k}_1) E^j(\omega_2,{\bf k}_2) \rangle
= - (4\pi)^2 \int \frac{d\omega_1}{2\pi}\frac{d\omega_2}{2\pi}
\\[2mm] \nonumber
&\times& 
\bigg[ \frac{1}{\omega_1} 
\frac{k^i_1k^k_1}{{\bf k}^2_1} + 
\frac{\omega_1}{\omega^2_1 - {\bf k}^2_1} \, 
\Big(\delta^{ik} - \frac{k^i_1k^k_1}{{\bf k}^2_1}\Big)
\bigg]
\bigg[\frac{1}{\omega_2} 
\frac{k^j_2k^l_2}{{\bf k}^2_2} + 
\frac{\omega_2}{\omega^2_2 - {\bf k}^2_2} \, 
\Big(\delta^{jl} - \frac{k^j_1k^l_1}{{\bf k}^2_1}\Big)
\bigg]
\\[2mm] \nonumber
&\times& \langle j^k(\omega_1,{\bf k}_1) j^j(\omega_2,{\bf k}_2) \rangle
\\[2mm] \nonumber
&=& - (4\pi e)^2 \int \frac{d\omega_1}{2\pi}\frac{d\omega_2}{2\pi}
\frac{d^3p_1}{(2\pi)^3}\frac{d^3p_2}{(2\pi)^3}\, v^k_1 v^l_2 
\\[2mm] \nonumber
&\times& 
\bigg[ \frac{1}{\omega_1} 
\frac{k^i_1k^k_1}{{\bf k}^2_1} + 
\frac{\omega_1}{\omega^2_1 - {\bf k}^2_1} \, 
\Big(\delta^{ik} - \frac{k^i_1k^k_1}{{\bf k}^2_1}\Big)
\bigg]
\bigg[ \frac{1}{\omega_2} 
\frac{k^j_2k^l_2}{{\bf k}^2_2} + 
\frac{\omega_2}{\omega^2_2 - {\bf k}^2_2} \, 
\Big(\delta^{jl} - \frac{k^j_1k^l_1}{{\bf k}^2_1}\Big)
\bigg]
\\[2mm] \nonumber
&\times& \langle \delta f(\omega_1,{\bf k}_1,{\bf p}_1) 
\delta f (\omega_2,{\bf k}_2,{\bf p}_2) \rangle \;.
\ea

As previously, we identify 
$\langle \delta f(\omega_1,{\bf k}_1,{\bf p}_1) 
\delta f (\omega_2,{\bf k}_2,{\bf p}_2) \rangle$
with $\langle \delta f(\omega_1,{\bf k}_1,{\bf p}_1) 
\delta f (\omega_2,{\bf k}_2,{\bf p}_2) \rangle_{\rm free}$
which equals
\ba
\langle \delta f(\omega_1,{\bf k}_1,{\bf p}_1) 
\delta f (\omega_2,{\bf k}_2,{\bf p}_2) \rangle_{\rm free}
&=& (2\pi )^3\delta^{(3)}({\bf p}_2 - {\bf p}_1) \:
2\pi \delta (\omega_1 + \omega_2) \:
(2\pi )^3 \delta^{(3)}({\bf k}_1 + {\bf k}_2) \: 
\\[2mm]\nonumber
&\times& 
2\pi\delta
\Big(\frac{\omega_1 - \omega_2}{2} 
- \frac{{\bf k}_1-{\bf k}_2}{2}{\bf v}_1\Big) \: 
f^0({\bf p}_1) \;.
\ea
Then, after performing trivial integrations,
$\langle E_0^i({\bf k}_1) E_0^j({\bf k}_2) \rangle$ equals
\be
\label{EE-0}
\langle E_0^i({\bf k}_1) E_0^j({\bf k}_2) \rangle
= -(4\pi e)^2 
(2\pi )^3 \delta^{(3)}({\bf k}_1 + {\bf k}_2) 
\int \frac{d^3p}{(2\pi)^3} \: f^0({\bf p}) \:
\frac{\big(({\bf k}_1 \cdot {\bf v})v^i - k^i_1\big)
\big(({\bf k}_2 \cdot {\bf v})v^j - k^j_2\big)}
{\big(({\bf k}_1 \cdot {\bf v})^2 - {\bf k}_1^2\big)
\big(({\bf k}_2 \cdot {\bf v})^2 - {\bf k}_2^2\big)}
\;.
\ee
Computing $k^i_1k^j_2\langle E_0^i({\bf k}_1) E_0^j({\bf k}_2) \rangle$,
one reproduces the result (\ref{ELEL-0}). Analogously to the correlation 
function $\langle E_0^i({\bf k}_1) E_0^j({\bf k}_2) \rangle$, 
one finds
\be
\langle E_0^i({\bf k}_1) 
\delta f_0({\bf k}_2,{\bf p}_2) \rangle
= 4\pi i e \: 
(2\pi )^3 \delta^{(3)}({\bf k}_1 + {\bf k}_2) \: 
f^0({\bf p}_2) \:
\frac{({\bf k}_1 \cdot {\bf v}_2)v^i_2 - k^i_1}
{({\bf k}_1 \cdot {\bf v}_2)^2 - {\bf k}_1^2} \;.
\ee

Starting with Eq.~(\ref{B-field-2}), we obtain
\be
\label{BB-0}
\langle B_0^i({\bf k}_1) B_0^j({\bf k}_2) \rangle
= -(4\pi e)^2 
(2\pi )^3 \delta^{(3)}({\bf k}_1 + {\bf k}_2) \:
\epsilon^{ikl} \epsilon^{jmn} k^k_1 k^m_2
\int \frac{d^3p}{(2\pi)^3} \: f^0({\bf p}) \:
\frac{v^l v^n}
{\big(({\bf k}_1 \cdot {\bf v})^2 - {\bf k}_1^2\big)
\big(({\bf k}_2 \cdot {\bf v})^2 - {\bf k}_2^2\big)}
\;,
\ee
and 
\be
\langle B_0^i({\bf k}_1) 
\delta f_0({\bf k}_2,{\bf p}_2) \rangle
= 4\pi i e \: 
(2\pi )^3 \delta^{(3)}({\bf k}_1 + {\bf k}_2) \: 
f^0({\bf p}_2) \:
\frac{\epsilon^{ijk}k^j_1 v^k_2}
{({\bf k}_1 \cdot {\bf v}_2)^2 - {\bf k}_1^2} \;.
\ee
Finally, one computes
\be
\label{EB-0}
\langle E_0^i({\bf k}_1) B_0^j({\bf k}_2) \rangle
= -(4\pi e)^2 
(2\pi )^3 \delta^{(3)}({\bf k}_1 + {\bf k}_2) 
\int \frac{d^3p}{(2\pi)^3} \: f^0({\bf p}) \:
\frac{\big(({\bf k}_1 \cdot {\bf v})v^i - k^i_1\big)
\epsilon^{jkl} k^k_2 v^l}
{\big(({\bf k}_1 \cdot {\bf v})^2 - {\bf k}_1^2\big)
\big(({\bf k}_2 \cdot {\bf v})^2 - {\bf k}_2^2\big)}
\;.
\ee


\section{Fluctuations of Longitudinal Electric Field
in Isotropic Plasma}
\label{sec-fluc-EL}


We first consider a special case of purely longitudinal electric
field in the isotropic plasma when the electric field is given by 
Eq.~(\ref{E-field-final-iso-long}). Then,
\ba
\label{E-fluc-iso-long}
k_1^i k_2^j \langle E^i(\omega_1,{\bf k}_1) 
E^j(\omega_2,{\bf k}_2) \rangle 
&= & \frac{1}
{\omega_1\omega_2 \varepsilon_L(\omega_1,{\bf k}_1)\,
\varepsilon_L(\omega_2,{\bf k}_2)}
\bigg[
\\[2mm] \nonumber
&\times& 16\pi^2 e^2
\int {d^3p_1 \over (2\pi)^3} \,
{d^3p_2 \over (2\pi)^3} 
\frac{{\bf k}_1\cdot {\bf v}_1}
{\omega_1 - {\bf k}_1 \cdot {\bf v}_1}\, 
\frac{{\bf k}_2\cdot {\bf v}_2}
{\omega_2 - {\bf k}_2 \cdot {\bf v}_2}
\langle \delta f_0({\bf k}_1,{\bf p}_1) 
\delta f_0({\bf k}_2,{\bf p}_2) \rangle
\\[2mm] \nonumber
&-& 4\pi i e 
\int {d^3p_1 \over (2\pi)^3} \,
\frac{{\bf k}_1\cdot {\bf v}_1}
{\omega_1 - {\bf k}_1 \cdot {\bf v}_1}\, 
k_2^j \langle \delta f_0({\bf k}_1,{\bf p}_1) 
E_0^j({\bf k}_2) \rangle
\\[2mm] \nonumber
&-& 4\pi i e 
\int {d^3p_2 \over (2\pi)^3} \, 
\frac{{\bf k}_2\cdot {\bf v}_2}
{\omega_2 - {\bf k}_2 \cdot {\bf v}_2}
k_1^i \langle E_0^i({\bf k}_1) \delta f_0({\bf k}_2,{\bf p}_2) \rangle
- k_1^ik_2^j \langle E_0^i({\bf k}_1) E_0^j({\bf k}_2) \rangle 
\bigg] \;.
\ea

Substituting the formulas of initial fluctuations
(\ref{ff-0}, \ref{fEL-0}, \ref{ELEL-0}) into 
Eq.~(\ref{E-fluc-iso-long}), one finds
\ba
\label{eq777}
k_1^i k_2^j \langle E^i(\omega_1,{\bf k}_1) 
E^j(\omega_2,{\bf k}_2) \rangle 
&=& 16\pi^2 e^2 \frac{(2\pi)^3\delta^{(3)}({\bf k}_2 + {\bf k}_1)}
{\omega_1\omega_2 \varepsilon_L(\omega_1,{\bf k}_1)\,
\varepsilon_L(\omega_2,{\bf k}_2)}
\\[2mm] \nonumber
&\times& 
\int {d^3p \over (2\pi)^3} \bigg[
\frac{{\bf k}_1\cdot {\bf v}}
{\omega_1 - {\bf k}_1 \cdot {\bf v}}\, 
\frac{{\bf k}_2\cdot {\bf v}}
{\omega_2 - {\bf k}_2 \cdot {\bf v}} 
+ \frac{{\bf k}_1\cdot {\bf v}}
{\omega_1 - {\bf k}_1 \cdot {\bf v}}
+ \frac{{\bf k}_2\cdot {\bf v}}
{\omega_2 - {\bf k}_2 \cdot {\bf v}} +1 \bigg] \, f^0({\bf p}) \;.
\ea
It appears that 
$$
\frac{{\bf k}_1\cdot {\bf v}}
{\omega_1 - {\bf k}_1 \cdot {\bf v}}\, 
\frac{{\bf k}_2\cdot {\bf v}}
{\omega_2 - {\bf k}_2 \cdot {\bf v}} 
+ \frac{{\bf k}_1\cdot {\bf v}}
{\omega_1 - {\bf k}_1 \cdot {\bf v}}
+ \frac{{\bf k}_2\cdot {\bf v}}
{\omega_2 - {\bf k}_2 \cdot {\bf v}} +1=
\frac{\omega_1}
{\omega_1 - {\bf k}_1 \cdot {\bf v}}\, 
\frac{\omega_2}
{\omega_2 - {\bf k}_2 \cdot {\bf v}}\;,
$$
and consequently Eq.~(\ref{eq777}) simplifies to
\ba
k_1^i k_2^j \langle E^i(\omega_1,{\bf k}_1) 
E^j(\omega_2,{\bf k}_2) \rangle 
= 16\pi^2 e^2 \frac{(2\pi)^3\delta^{(3)}({\bf k}_2 + {\bf k}_1)}
{\varepsilon_L(\omega_1,{\bf k}_1)\,
\varepsilon_L(\omega_2,{\bf k}_2)}
\int {d^3p \over (2\pi)^3} 
\frac{f^0({\bf p})}
{(\omega_1 - {\bf k}_1 \cdot {\bf v})
(\omega_2 - {\bf k}_2 \cdot {\bf v})} \, \;.
\ea
This equation could be easily obtained directly from 
Eq.~(\ref{E-field-final-iso-long2}) where the initial electric
field is already eliminated. 

Taking into account that the electric 
fields are parallel to their wave vectors, and consequently 
$k_1^i k_1^j \langle E^i(\omega_1,{\bf k}_1) 
E^j(\omega_2,-{\bf k}_1) \rangle = {\bf k}_1^2
\langle E^i(\omega_1,{\bf k}_1) E^i(\omega_2,-{\bf k}_1) \rangle$,
one finally finds
\ba
\label{E-fluc-iso-long-2}
\langle E^i(\omega_1,{\bf k}_1) 
E^i(\omega_2,{\bf k}_2) \rangle 
= - 16\pi^2 e^2 \frac{(2\pi)^3\delta^{(3)}({\bf k}_2 + {\bf k}_1)}
{{\bf k}_1^2 \varepsilon_L(\omega_1,{\bf k}_1)\,
\varepsilon_L(\omega_2,-{\bf k}_1)}
\int {d^3p \over (2\pi)^3} 
\frac{f^0({\bf p})}
{(\omega_1 - {\bf k}_1 \cdot {\bf v})
(\omega_2 + {\bf k}_1 \cdot {\bf v})} \, \;.
\ea

Let us now compute 
$\langle E^i(t_1,{\bf r}_1)E^i(t_2,{\bf r}_2) \rangle$ given by 
\ba
\label{EL-fluc-x}
\langle E^i(t_1,{\bf r}_1) E^i(t_2,{\bf r}_2) \rangle
&=& \int_{-\infty +i\sigma}^{\infty +i\sigma}
{d\omega_1 \over 2\pi}
\int_{-\infty +i\sigma}^{\infty +i\sigma}
{d\omega_2 \over 2\pi}
\int {d^3k_1 \over (2\pi)^3}
\int {d^3k_2 \over (2\pi)^3}
\\[2mm] \nonumber
&\times& e^{-i(\omega_1 t_1 - {\bf k}_1\cdot {\bf r}_1
+ \omega_2 t_2 - {\bf k}_2\cdot {\bf r}_2)}
\langle E^i(\omega_1,{\bf k}_1)
E^i(\omega_2,{\bf k}_2) \rangle \;.
\ea
Zeros of $\varepsilon_L(\omega_i,{\bf k}_i)$ and 
of the denominators $(\omega_i - {\bf k}_i \cdot {\bf v} +i0^+)$ 
with $i=1,2$ contribute to the integrals over $\omega_1$ and 
$\omega_2$. However, once the plasma system under consideration 
is stable with respect to longitudinal modes, all zeros of 
$\varepsilon_L$ lie in the lower half-plane of complex $\omega$.
Consequently, the contributions associated with these zeros
exponentially decay in time and they vanish in the long time 
limit of both $t_1$ and $t_2$. 

We are further interested in the long time limit of 
$\langle E^i(t_1,{\bf r}_1) E^i(t_2,{\bf r}_2) \rangle$ and 
then, the only non-vanishing contribution corresponds to the 
poles at $\omega_1 = {\bf k}_1 \cdot {\bf v}$ and 
$\omega_2 = {\bf k}_2 \cdot {\bf v}$. This contribution reads
\be
\langle E^i(t_1,{\bf r}_1) E^i(t_2,{\bf r}_2) \rangle_\infty
= 16\pi^2 e^2 
\int {d^3k \over (2\pi)^3} 
\int {d^3p \over (2\pi)^3}
e^{-i{\bf k} \cdot 
\big({\bf v}(t_1 - t_2) - ({\bf r}_1 - {\bf r}_2)\big)}
\frac{ f^0({\bf p})}{{\bf k}^2 
\varepsilon_L({\bf k} \cdot {\bf v},{\bf k})\,
\varepsilon_L(-{\bf k} \cdot {\bf v},-{\bf k})} \;.
\ee
Keeping in mind that 
$\varepsilon_L(-\omega,-{\bf k})= \varepsilon_L^*(\omega,{\bf k})$ 
for real $\omega$ and ${\bf k}$, it can be rewritten as
\be 
\label{EE-tx-final}
\langle E^i(t_1,{\bf r}_1) E^i(t_2,{\bf r}_2) \rangle_\infty 
= 32\pi^3 e^2 
\int {d\omega \over 2\pi}
{d^3k \over (2\pi)^3} 
\frac{e^{-i \big(\omega(t_1 - t_2) - 
{\bf k} \cdot ({\bf r}_1 - {\bf r}_2)\big)}}
{{\bf k}^2 
|\varepsilon_L(\omega,{\bf k})|^2}
\int {d^3p \over (2\pi)^3} \:
\delta(\omega - {\bf k}\cdot {\bf v})
f^0({\bf p}) \;.
\ee
As seen, $\langle E^i(t_1,{\bf r}_1) E^i(t_2,{\bf r}_2) \rangle_\infty$
given by Eq.~(\ref{EE-tx-final}) depends on $t_1, t_2$ and 
${\bf r}_1, {\bf r}_2$ only through $(t_1 - t_2)$ and 
$({\bf r}_1 - {\bf r}_2)$ and it can be written as
\be
\langle E^i(t_1,{\bf r}_1) E^i(t_2,{\bf r}_2) \rangle_\infty
= \int {d\omega \over 2\pi}
{d^3k \over (2\pi)^3} 
e^{-i \big(\omega(t_1 - t_2) - 
{\bf k} \cdot ({\bf r}_1 - {\bf r}_2)\big)}
\langle E^i E^i\rangle_{\omega {\bf k}} \;,
\ee
where the fluctuation spectrum 
$\langle E^i E^i\rangle_{\omega {\bf k}}$ is
\be
\label{EE-spectrum}
\langle E^i E^i\rangle_{\omega {\bf k}} 
= \frac{32\pi^3 e^2}{{\bf k}^2 
|\varepsilon_L(\omega,{\bf k})|^2}
\int {d^3p \over (2\pi)^3} \:
\delta(\omega - {\bf k}\cdot {\bf v})
f^0({\bf p}) \;.
\ee

In the case of equilibrium plasma, the formula (\ref{EE-spectrum}) 
provides the result which can be obtained directly by means of the 
fluctuation-dissipation theorem. Let us derive the result.
Due to the identity
\be
\label{iden1}
\frac{1}{x\pm i0^+} = {\cal P}\frac{1}{x} \mp i\pi \delta (x) \;,
\ee
the imaginary part of $\varepsilon_L(\omega,{\bf k})$, which is 
given by Eq.~(\ref{eL}), equals 
\be
\Im \varepsilon_L(\omega,{\bf k}) =
- \frac{4\pi^2 e^2}{{\bf k}^2} 
\int {d^3p \over (2\pi)^3} \:
\delta(\omega - {\bf k}\cdot {\bf v}) \:
{\bf k} \cdot \frac{\partial f^0({\bf p})}{\partial{\bf p}} \;.
\ee
In equilibrium $f^0({\bf p}) \sim e^{-\beta E_p}$ and 
$\partial f^0({\bf p})/\partial{\bf p} = - \beta {\bf v} f^0({\bf p})$.
Therefore, $\Im \varepsilon_L$ equals
\be
\label{Im-eL-eq}
\Im \varepsilon_L(\omega,{\bf k}) =
\frac{4\pi^2 e^2}{T{\bf k}^2} 
\int {d^3p \over (2\pi)^3} \:
\delta(\omega - {\bf k}\cdot {\bf v}) 
\:({\bf k} \cdot {\bf v}) \: f^0({\bf p})
= \frac{4\pi^2 e^2 \omega}{T{\bf k}^2} 
\int {d^3p \over (2\pi)^3} \:
\delta(\omega - {\bf k}\cdot {\bf v}) \: f^0({\bf p})
\;.
\ee

Using the expression (\ref{Im-eL-eq}), the formula (\ref{EE-spectrum})
is rewritten as
\be
\label{EE-spectrum-eq}
\langle E^i E^i\rangle_{\omega {\bf k}} 
= 8\pi \frac{T}{\omega}
\frac{\Im \varepsilon_L(\omega,{\bf k})} 
{|\varepsilon_L(\omega,{\bf k})|^2}
\;,
\ee
which agrees with Eq.~(51.25) from \cite{LP81} which is obtained
there in essentially the same way.


\section{Fluctuations of Magnetic Field in Isotropic Plasma}
\label{sec-fluc-B}


As seen in Eq.~(\ref{B-field-final-iso}), the magnetic field in 
isotropic plasma is given by three terms. Therefore, nine terms 
enter the correlation function 
$\langle B^i(\omega_1,{\bf k}_1) B^j(\omega_2,{\bf k}_2)\rangle$.
Substituting into these terms the initial fluctuations derived in
Sec.~\ref{sec-initial-fluc}, one finds after an elementary but 
lengthy and tedious analysis the following expression
\ba
\label{B-fluc-iso3}
\langle B^i(\omega_1,{\bf k}_1) B^j(\omega_2,{\bf k}_2)\rangle 
&=&\frac{(4\pi e)^2 (2\pi )^3\delta^{(3)}({\bf k}_1 + {\bf k}_2)
\: \epsilon^{ikl}\epsilon^{jmn} k_1^k k_2^m }
{\big(\omega_1^2 \varepsilon_T(\omega_1,{\bf k}_1)-{\bf k}_1^2)\big)
\big(\omega_2^2 \varepsilon_T(\omega_2,{\bf k}_2)-{\bf k}_2^2)\big)}
\\[2mm] \nonumber
&\times& 
\int {d^3p \over (2\pi)^3}\, f^0({\bf p})
\frac{v^l v^n}{
(\omega_1 - {\bf k}_1\cdot {\bf v})
(\omega_2 - {\bf k}_2\cdot {\bf v})
(({\bf k}_1\cdot {\bf v})^2 -{\bf k}_1^2)
(({\bf k}_2\cdot {\bf v})^2 -{\bf k}_2^2)}
\\[2mm] \nonumber
&\times& 
\Big[
(\omega_1 ({\bf k}_1\cdot {\bf v}) - {\bf k}_1^2)
+ \omega_1 \varepsilon_T(\omega_1,{\bf k}_1)
(\omega_1 - {\bf k}_1\cdot {\bf v}) 
\Big]
\\[2mm] \nonumber
&\times&
\Big[
(\omega_2 ({\bf k}_2\cdot {\bf v}) - {\bf k}_2^2)
+ \omega_2 \varepsilon_T(\omega_2,{\bf k}_2)
(\omega_2 - {\bf k}_2\cdot {\bf v}) 
\Big] \;.
\ea

We now compute 
$\langle B^i(t_1,{\bf r}_1)B^j(t_2,{\bf r}_2) \rangle$ given by 
\ba
\langle B^i(t_1,{\bf r}_1) B^j(t_2,{\bf r}_2) \rangle
&=& \int_{-\infty +i\sigma}^{\infty +i\sigma}
{d\omega_1 \over 2\pi}
\int_{-\infty +i\sigma}^{\infty +i\sigma}
{d\omega_2 \over 2\pi}
\int {d^3k_1 \over (2\pi)^3}
\int {d^3k_2 \over (2\pi)^3}
\\[2mm] \nonumber
&\times& e^{-i(\omega_1 t_1 - {\bf k}_1\cdot {\bf r}_1
+ \omega_2 t_2 - {\bf k}_2\cdot {\bf r}_2)}
\langle B^i(\omega_1,{\bf k}_1)
B^j(\omega_2,{\bf k}_2) \rangle \;.
\ea
Zeros of $(\omega_i^2\varepsilon_T(\omega_i,{\bf k}_i)- {\bf k}_i^2)$ 
and of $\omega_i - {\bf k}_i \cdot {\bf v} +i0^+)$ 
with $i=1,2$ contribute to the integrals over $\omega_1$ and 
$\omega_2$. However, once the plasma system under consideration 
is stable with respect to transverse modes, all zeros of 
$(\omega_i^2\varepsilon_T(\omega_i,{\bf k}_i)- {\bf k}_i^2)$
lie in the lower half-plane of complex $\omega$.
Consequently, the contributions associated with these zeros
exponentially decay in time and they vanish in the long time 
limit of both $t_1$ and $t_2$. 

We further consider the long time limit of 
$\langle B^i(t_1,{\bf r}_1) B^j(t_2,{\bf r}_2) \rangle$ and 
then, the only non-vanishing contribution corresponds to the 
poles at $\omega_1 = {\bf k}_1 \cdot {\bf v}$ and 
$\omega_2 = {\bf k}_2 \cdot {\bf v}$. This contribution reads
\ba
\langle B^i(t_1,{\bf r}_1) B^j(t_2,{\bf r}_2) \rangle_\infty
&=& - \int 
{d^3k_1 \over (2\pi)^3}
{d^3k_2 \over (2\pi)^3}
{d^3p \over (2\pi)^3}\, f^0({\bf p}) \;
e^{-i(\omega_1 t_1 - {\bf k}_1\cdot {\bf r}_1
+ \omega_2 t_2 - {\bf k}_2\cdot {\bf r}_2)}
\\[2mm] \nonumber
&\times& 
\frac{(4\pi e)^2 (2\pi )^3\delta^{(3)}({\bf k}_1 + {\bf k}_2)
\: \epsilon^{ikl}\epsilon^{jmn} k_1^k k_2^m }
{\big(\omega_1^2 \varepsilon_T(\omega_1,{\bf k}_1)-{\bf k}_1^2)\big)
\big(\omega_2^2 \varepsilon_T(\omega_2,{\bf k}_2)-{\bf k}_2^2)\big)}
\\[2mm] \nonumber
&\times& 
\frac{v^l v^n}{
(({\bf k}_1\cdot {\bf v})^2 -{\bf k}_1^2)
(({\bf k}_2\cdot {\bf v})^2 -{\bf k}_2^2)}
(\omega_1 ({\bf k}_1\cdot {\bf v}) - {\bf k}_1^2)
(\omega_2 ({\bf k}_2\cdot {\bf v}) - {\bf k}_2^2)
\Bigg|_{\omega_1 = {\bf k}_1 \cdot {\bf v},\;\;\;
\omega_2 = {\bf k}_2 \cdot {\bf v}}
\;.
\ea
It can be easily expressed as 
\ba
\langle B^i(t_1,{\bf r}_1) B^j(t_2,{\bf r}_2) \rangle_\infty 
&=& 
\int {d\omega \over 2\pi} {d^3k \over (2\pi)^3}
e^{-i \big(\omega (t_1 - t_2)
 - {\bf k}\cdot ({\bf r}_1 - {\bf r}_2)\big)}
\langle B^i B^j\rangle_{\omega \, {\bf k}} \;,
\ea
where the fluctuation spectrum is
\ba
\langle B^i B^j\rangle_{\omega \, {\bf k}}
=\frac{32 \pi^3 e^2 \epsilon^{ikl}\epsilon^{jmn} k^k k^m}
{\big(\omega^2 \varepsilon_T(\omega,{\bf k})-{\bf k}^2)\big)
\big(\omega^2 \varepsilon_T(-\omega,-{\bf k})-{\bf k}^2)\big)}
\int {d^3p \over (2\pi)^3}\, f^0({\bf p}) \:
\delta (\omega - {\bf k} \cdot {\bf v}) \: v^l v^n \;.
\ea
When both $\omega$ and ${\bf k}$ are real 
$\varepsilon_T(-\omega,-{\bf k}) = \varepsilon_T^*(\omega,{\bf k})$.
Therefore, the fluctuation spectrum can be rewritten as
\ba
\label{BiBj-spec}
\langle B^i B^j\rangle_{\omega \, {\bf k}}
=\frac{32 \pi^3 e^2 \epsilon^{ikl}\epsilon^{jmn} k^k k^m}
{\big|\omega^2 \varepsilon_T(\omega,{\bf k})-{\bf k}^2\big|^2}
\int {d^3p \over (2\pi)^3}\, f^0({\bf p}) \:
\delta (\omega - {\bf k} \cdot {\bf v}) \: v^l v^n
\;.
\ea

One observes that the matrix function 
\be
\label{M-def}
M^{ij}(\omega,{\bf k}) \equiv
\int {d^3p \over (2\pi)^3}\, f^0({\bf p}) \:
\delta (\omega - {\bf k} \cdot {\bf v}) \: v^i v^j \;,
\ee
which enters the correlation function (\ref{BiBj-spec}), can 
be decomposed as
\be
\label{M-L-T}M^{ij}(\omega,{\bf k}) = 
M_L(\omega,{\bf k}) \: \frac{k^ik^j}{{\bf k}^2} + 
M_T(\omega,{\bf k}) \:
\Big(\delta^{ij} - \frac{k^ik^j}{{\bf k}^2}\Big) \;,
\ee
because the plasma is assumed to be isotropic. Comparing 
Eq.~(\ref{M-L-T}) to Eq.~(\ref{M-def}), one finds
\be
\label{M_L}
M_L(\omega,{\bf k}) \equiv
\int {d^3p \over (2\pi)^3}\, f^0({\bf p}) \:
\delta (\omega - {\bf k} \cdot {\bf v}) \: 
\frac{({\bf k}\cdot {\bf v})^2}{{\bf k}^2}
\;,
\ee\be
\label{M_T}
M_T(\omega,{\bf k}) \equiv
\frac{1}{2}
\int {d^3p \over (2\pi)^3}\, f^0({\bf p}) \:
\delta (\omega - {\bf k} \cdot {\bf v}) \: 
\bigg[{\bf v}^2 -
\frac{({\bf k}\cdot {\bf v})^2}{{\bf k}^2}
\bigg] \;.
\ee
Using the decomposition (\ref{M-L-T}), the correlation function 
(\ref{BiBj-spec}) can be written down as
\ba
\label{BiBj-M-spec}
\langle B^i B^j\rangle_{\omega \, {\bf k}}
=\frac{32 \pi^3 e^2 (\delta^{ij}{\bf k}^2 - k^i k^j )}
{\big|\omega^2 \varepsilon_T(\omega,{\bf k})-{\bf k}^2\big|^2}
M_T(\omega,{\bf k})\;.
\ea

For equilibrium plasma the correlation function 
$\langle B^i B^j\rangle_{\omega \, {\bf k}}$ can be expressed 
in the form of fluctuation-dissipation relation. One first 
observes that due to the identity (\ref{iden1}), the imaginary 
part of $\varepsilon_T(\omega,{\bf k})$, which is given by 
Eq.~(\ref{eT}), is 
\be
\Im \varepsilon_T(\omega,{\bf k}) =
- \frac{2\pi^2 e^2}{\omega} 
\int {d^3p \over (2\pi)^3} \:
\delta(\omega - {\bf k}\cdot {\bf v}) \:
\bigg[
{\bf v} \cdot \frac{\partial f^0({\bf p})}{\partial{\bf p}} 
- \frac{{\bf k} \cdot {\bf v}}{{\bf k}^2} \;
{\bf k} \cdot \frac{\partial f^0({\bf p})}{\partial{\bf p}} 
\bigg]\;. 
\ee
With the equilibrium distribution function, $\Im \varepsilon_T$ equals
\be
\label{Im-eT-eq}
\Im \varepsilon_T(\omega,{\bf k}) =
\frac{2\pi^2 e^2}{T \omega \, {\bf k}^2} \int {d^3p \over (2\pi)^3} \:
\delta(\omega - {\bf k}\cdot {\bf v}) \:
\big({\bf k}^2{\bf v}^2 
- ({\bf k} \cdot {\bf v})^2\big)\: f^0({\bf p})
\;.
\ee
Consequently, the function $M_T$ (\ref{M_T}) can be expressed 
through $\Im \varepsilon_T$ (\ref{Im-eT-eq}) as
\be
M_T(\omega,{\bf k}) = \frac{T\omega}{4\pi^2 e^2} \:
\Im \varepsilon_T(\omega,{\bf k}) \;,
\ee 
and finally,
\ba
\label{BiBj-spec-eq}
\langle B^i B^j\rangle_{\omega \, {\bf k}}
=\frac{8 \pi \, T}{\omega^3}
\: (\delta^{ij}{\bf k}^2 - k^i k^j ) \:
\frac{\Im \varepsilon_T (\omega,{\bf k})}
{\big|\varepsilon_T(\omega,{\bf k})-
\frac{{\bf k}^2}{\omega^2}\big|^2} \;.
\ea
Eq.~(\ref{BiBj-spec-eq}) coincides with the formula (11.2.2.7) 
from \cite{Akh75} obtained there directly from the 
fluctuation-dissipation theorem.

                                                                                
\section{Fluctuations of Electric Field in Isotropic Plasma}
\label{sec-fluc-E}
                                                                                

The analysis of electric field fluctuations is much more complicated 
than that of the magnetic field. First of all, there are five terms 
which enter the formula of electric field given by 
Eq.~(\ref{E-field-final-iso}), and consequently, the correlation function 
$\langle E^i(\omega_1,{\bf k}_1) E^j(\omega_2,{\bf k}_2) \rangle$ 
includes 25 terms. The magnetic field is purely transverse 
and some terms automatically drop out but the electric fields have 
longitudinal and transverse components. Using the formulas of 
initial fluctuations, which are derived in Sec.~\ref{sec-initial-fluc},  
and patiently analyzing term by term, one obtains after an elementary 
but very lengthy calculation the correlation function of the form 
\ba
\langle E^i(\omega_1,{\bf k}_1)  E^j(\omega_2,{\bf k}_2) \rangle
&=& 16 \pi^2 e^2 (2\pi )^3\delta^{(3)}({\bf k}_1 + {\bf k}_2)
\int {d^3p \over (2\pi)^3} \, f^0({\bf p})
\\[2mm]\nonumber
&\times \Bigg\{&
\frac{k_1^i}{\omega_1^2 \varepsilon_L(\omega_1,{\bf k}_1)} \:
\frac{k_2^j}{\omega_2^2 \varepsilon_L(\omega_2,{\bf k}_2)} \:
\frac{\omega_1^2 \omega_2^2}
{{\bf k}_1^2 
(\omega_1 - {\bf k}_1\cdot {\bf v}) \:
{\bf k}_2^2 
(\omega_2 - {\bf k}_2\cdot {\bf v})}
\\[2mm]\nonumber
&+&
\frac{k_1^i}{\omega_1^2 \varepsilon_L(\omega_1,{\bf k}_1)} \:
\frac{ v^j {\bf k}_2^2 - k_2^j ({\bf k}_2 \cdot {\bf v})}
{\omega_2^2 \varepsilon_T(\omega_2,{\bf k}_2)-{\bf k}_2^2}\:
\frac{
\omega_1^2 
[\omega_2
(\omega_2({\bf k}_2\cdot {\bf v}) - {\bf k}_2^2)
-
{\bf k}_2^2 
(\omega_2 - {\bf k}_2\cdot {\bf v})]}
{{\bf k}_1^2
(\omega_1 - {\bf k}_1\cdot {\bf v})\:
{\bf k}_2^2
(\omega_2 - {\bf k}_2\cdot {\bf v})\:
(({\bf k}_2\cdot {\bf v})^2 - {\bf k}_2^2) }
\\[2mm]\nonumber
&+&
\frac{ v^i {\bf k}_1^2 - k_1^i ({\bf k}_1 \cdot {\bf v})}
{\omega_1^2 \varepsilon_T(\omega_1,{\bf k}_1)-{\bf k}_1^2}\:
\frac{k_2^j}{\omega_2^2 \varepsilon_L(\omega_2,{\bf k}_2)} \:
\frac{
\omega_2^2 
[\omega_1 
(\omega_1 ({\bf k}_1\cdot {\bf v}) - {\bf k}_1^2)
-
{\bf k}_1^2
(\omega_1 - {\bf k}_1\cdot {\bf v})]}
{{\bf k}_1^2
(\omega_1 - {\bf k}_1\cdot {\bf v})\:
(({\bf k}_1\cdot {\bf v})^2 - {\bf k}_1^2)\:
{\bf k}_2^2
(\omega_2 - {\bf k}_2\cdot {\bf v})}
\\[2mm]\nonumber
&+&
\frac{k_1^i ({\bf k}_1\cdot {\bf v}) - v^i{\bf k}_1^2}
{\omega_1^2 \varepsilon_T(\omega_1,{\bf k}_1)-{\bf k}_1^2} \:
\frac{k_2^j ({\bf k}_2\cdot {\bf v}) - v^j{\bf k}_2^2}
{\omega_2^2 \varepsilon_T(\omega_2,{\bf k}_2)-{\bf k}_2^2}
\\[2mm]\nonumber
&&\times
\frac{
\omega_1 (\omega_1 ({\bf k}_1\cdot {\bf v}) - {\bf k}_1^2)
-
{\bf k}_1^2 (\omega_1 - {\bf k}_1\cdot {\bf v})}
{{\bf k}_1^2
(\omega_1 - {\bf k}_1\cdot {\bf v})\:
(({\bf k}_1\cdot {\bf v})^2 - {\bf k}_1^2)}
\;
\frac{
\omega_2 (\omega_2 ({\bf k}_2\cdot {\bf v}) - {\bf k}_2^2)
-
{\bf k}_2^2 (\omega_2 - {\bf k}_2\cdot {\bf v})}
{{\bf k}_2^2
(\omega_2 - {\bf k}_2\cdot {\bf v})\:
(({\bf k}_2\cdot {\bf v})^2 - {\bf k}_2^2)}
\Bigg\} \;.
\ea

We now compute 
$\langle E^i(t_1,{\bf r}_1)E^j(t_2,{\bf r}_2) \rangle$ given by 
\ba
\langle E^i(t_1,{\bf r}_1) E^j(t_2,{\bf r}_2) \rangle
&=& \int_{-\infty +i\sigma}^{\infty +i\sigma}
{d\omega_1 \over 2\pi}
\int_{-\infty +i\sigma}^{\infty +i\sigma}
{d\omega_2 \over 2\pi}
\int {d^3k_1 \over (2\pi)^3}
\int {d^3k_2 \over (2\pi)^3}
\\[2mm] \nonumber
&\times& e^{-i(\omega_1 t_1 - {\bf k}_1\cdot {\bf r}_1
+ \omega_2 t_2 - {\bf k}_2\cdot {\bf r}_2)}
\langle E^i(\omega_1,{\bf k}_1)
E^j(\omega_2,{\bf k}_2) \rangle \;.
\ea
Zeros of $(\omega_i^2\varepsilon_T(\omega_i,{\bf k}_i)- {\bf k}_i^2)$,
$\omega_i^2\varepsilon_L(\omega_i,{\bf k}_i)$ and of 
$(\omega_i - {\bf k}_i \cdot {\bf v} +i0^+)$ with $i=1,2$ 
contribute to the integrals over $\omega_1$ and $\omega_2$. 
However, once the plasma system under consideration is stable, all 
zeros of $(\omega_i^2\varepsilon_T(\omega_i,{\bf k}_i)- {\bf k}_i^2)$
and $\omega_i^2\varepsilon_L(\omega_i,{\bf k}_i)$ lie in the lower 
half-plane of complex $\omega$. Consequently, the contributions 
associated with these zeros exponentially decay in time and they 
vanish in the long time limit of both $t_1$ and $t_2$. 

We further consider the long time limit of 
$\langle E^i(t_1,{\bf r}_1) E^j(t_2,{\bf r}_2) \rangle$ and 
then, the only non-vanishing contribution corresponds to the 
poles at $\omega_1 = {\bf k}_1 \cdot {\bf v}$ and 
$\omega_2 = {\bf k}_2 \cdot {\bf v}$. This contribution reads
\ba
\label{EiEj-tr-1}
&&\langle E^i(t_1,{\bf r}_1) 
E^j(t_2,{\bf r}_2) \rangle_\infty
= - 16 \pi^2 e^2 
\int {d^3k_1 \over (2\pi)^3} {d^3k_2 \over (2\pi)^3} \:
(2\pi )^3\delta^{(3)}({\bf k}_1 + {\bf k}_2)
\\[2mm] \nonumber 
&\times&
\int {d^3p \over (2\pi)^3} \, f^0({\bf p}) \:
e^{-i(\omega_1 t_1 - {\bf k}_1\cdot {\bf r}_1
+ \omega_2 t_2 - {\bf k}_2\cdot {\bf r}_2)}
\frac{\omega_1 \omega_2}
{{{\bf k}_1^2 {\bf k}_2^2}}
\\[2mm] \nonumber
&\times \Bigg[&
\frac{\omega_1 k_1^i}{\omega_1^2 \varepsilon_L(\omega_1,{\bf k}_1)} 
+
\frac{k_1^i ({\bf k}_1\cdot {\bf v}) - v^i{\bf k}_1^2}
{\omega_1^2 \varepsilon_T(\omega_1,{\bf k}_1)-{\bf k}_1^2} 
\Bigg] 
\Bigg[
\frac{\omega_2 k_2^j}{\omega_2^2 \varepsilon_L(\omega_2,{\bf k}_2)} 
+
\frac{ v^j {\bf k}_2^2 - k_2^j ({\bf k}_2 \cdot {\bf v})}
{\omega_2^2 \varepsilon_T(\omega_2,{\bf k}_2)-{\bf k}_2^2} 
\Bigg]
\Bigg|_{\omega_1={\bf k}_1\cdot {\bf v}, \;,\;
\omega_2={\bf k}_2\cdot {\bf v}} \;.
\ea

The correlation function (\ref{EiEj-tr-1}) can be rewritten as
\ba
\langle E^i(t_1,{\bf r}_1) E^j(t_2,{\bf r}_2) \rangle_\infty 
&=& 
\int {d\omega \over 2\pi} {d^3k \over (2\pi)^3}
e^{-i \big(\omega (t_1 - t_2)
 - {\bf k}\cdot ({\bf r}_1 - {\bf r}_2)\big)}
\langle E^i E^j\rangle_{\omega \, {\bf k}} \;,
\ea
where the fluctuation spectrum is
\ba
\label{EiEj-spec-1}
&&\langle E^i E^j\rangle_{\omega \, {\bf k}}
= 16 \pi^2 e^2 
\int {d^3p \over (2\pi)^3} \, f^0({\bf p}) \:
2\pi \delta (\omega - {\bf k} \cdot {\bf v})
\frac{\omega^2}{{\bf k}^4}
\\[2mm] \nonumber 
&\times \Bigg\{&
\frac{k^i}{\omega^2 \varepsilon_L(\omega,{\bf k})} \:
\frac{k^j}{\omega^2 \varepsilon_L(-\omega,{\bf k}^2)} \:
\omega^2
+
\frac{k^i}{\omega^2 \varepsilon_L(\omega,{\bf k})} \:
\frac{ v^j {\bf k}^2 - k^j ({\bf k} \cdot {\bf v})}
{\omega^2 \varepsilon_T(-\omega,{\bf k})-{\bf k}^2}\: \omega
\\[2mm]\nonumber
&+&
\frac{ v^i {\bf k}^2 - k^i ({\bf k} \cdot {\bf v})}
{\omega^2 \varepsilon_T(\omega,{\bf k})-{\bf k}^2}\:
\frac{k^j}{\omega^2 \varepsilon_L(-\omega,{\bf k})} \:
\omega
+
\frac{k^i ({\bf k} \cdot {\bf v}) - v^i{\bf k}^2}
{\omega^2 \varepsilon_T(\omega,{\bf k})-{\bf k}^2} \:
\frac{k^j ({\bf k}\cdot {\bf v}) - v^j{\bf k}^2}
{\omega^2 \varepsilon_T(-\omega,{\bf k})-{\bf k}^2}
\Bigg\} 
\;.
\ea
One easily proves that the second and third contribution to the 
fluctuation spectrum (\ref{EiEj-spec-1}) vanish due to the plasma 
isotropy. Taking into account that for real $\omega$ and ${\bf k}$,
the dielectric function obeys
$\varepsilon_s(-\omega,-{\bf k})= \varepsilon_s^*(\omega,{\bf k})$
with $s=L,T$, the fluctuation spectrum (\ref{EiEj-spec-1}) can be 
written as 
\ba
\label{EiEj-spec-2}
\langle E^i E^j\rangle_{\omega \, {\bf k}}
&=& 16 \pi^2 e^2 
\int {d^3p \over (2\pi)^3} \, f^0({\bf p}) \:
2\pi \delta (\omega - {\bf k} \cdot {\bf v})
\frac{\omega^2}{{\bf k}^4}
\\[2mm] \nonumber 
&\times \Bigg\{&
\frac{ \omega^2 k^ik^j}{|\omega^2 \varepsilon_L(\omega,{\bf k})|^2}
+
\frac{\big( k^i ({\bf k} \cdot {\bf v}) - v^i{\bf k}^2 \big)
\big(k^j ({\bf k}\cdot {\bf v}) - v^j{\bf k}^2 \big)}
{|\omega^2 \varepsilon_T(\omega,{\bf k})-{\bf k}^2|^2}
\Bigg\} 
\;.
\ea
Due to the plasma isotropy, the expression, which enters the 
transverse contribution, can be further rewritten as 
\ba
\int {d^3p \over (2\pi)^3} \, f^0({\bf p}) \:
2\pi \delta (\omega - {\bf k} \cdot {\bf v})
\big( k^i ({\bf k} \cdot {\bf v}) - v^i{\bf k}^2 \big)
\big(k^j ({\bf k}\cdot {\bf v}) - v^j{\bf k}^2 \big)
\\[2mm] \nonumber 
= \frac{1}{2}
\Big(\delta^{ij} - \frac{k^ik^j}{{\bf k}^2}\Big)
{\bf k}^2
\int {d^3p \over (2\pi)^3} \, f^0({\bf p}) \:
2\pi \delta (\omega - {\bf k} \cdot {\bf v}) 
\big(({\bf k}^2 {\bf v}^2 - ({\bf k}\cdot {\bf v})^2 \big) \;.
\ea

In the equilibrium plasma, the imaginary parts of 
$\varepsilon_L(\omega,{\bf k})$ and $\varepsilon_T(\omega,{\bf k})$
are given by the formulas (\ref{Im-eL-eq}, \ref{Im-eT-eq}) and
the fluctuation spectrum (\ref{EiEj-spec-2}) can be 
expressed as 
\ba
\label{EiEj-spec-FD}
\langle E^i E^j\rangle_{\omega \, {\bf k}}
= 
8\pi T \omega^3 \bigg[
\frac{k^ik^j}{{\bf k}^2}
\frac{\Im \varepsilon_L(\omega,{\bf k})}
{|\omega^2 \varepsilon_L(\omega,{\bf k})|^2}
+
\Big(\delta^{ij} - \frac{k^ik^j}{{\bf k}^2}\Big)
\frac{\Im \varepsilon_T(\omega,{\bf k})}
{|\omega^2 \varepsilon_T(\omega,{\bf k})-{\bf k}^2|^2}
\bigg] \;,
\ea
which for the longitudinal fields reproduces the formula 
(\ref{EE-spectrum-eq}). The result (\ref{EiEj-spec-FD}) 
agrees with Eq.~(11.2.2.6) from \cite{Akh75} derived using 
the fluctuation-dissipation relation.


\section{Fluctuations of longitudinal electric field in 
the two-stream system}
\label{sec-2-streams}


Nonequlibrium calculations are usually much more difficult than 
the equilibrium ones. The first problem is to invert the matrix
$\Sigma^{ij}(\omega,{\bf k})$ defined by Eq.~(\ref{matrix-sigma}).
In the case of longitudinal electric field, which is discussed
here, it is solved trivially. We start with Eq.~(\ref{E-field2})
projecting it on ${\bf k}$ and assuming that ${\bf E}$ and ${\bf E}_0$ 
are purely longitudinal. Then, the matrix (\ref{matrix-sigma}) is
replaced by the scalar function. 

Further, we neglect the first term in the r.h.s. of Eq.~(\ref{E-field2}). 
This term vanishes in isotropic systems; it is of order $e^2$ higher 
than the second term; it is also expected to be small in nonrelativistic 
regime due to the smallness of particle velocity. So, there are good 
reasons to neglect it. Eliminating ${\bf E}_0$ by means of the first 
Maxwell equation we obtain Eq.~(\ref{E-field-final-iso-long2}) which 
was previously derived for the case of isotropic plasma. In the
following we consider fluctuations of longitudinal electric fields 
in the nonrelativistic two-stream system.

The distribution function of the two-stream system is chosen to be 
\be
\label{f-2-streams}
f^0({\bf p}) = (2\pi )^3 n 
\Big[\delta^{(3)}({\bf p} - {\bf q})
+ \delta^{(3)}({\bf p} + {\bf q}) \Big] \;,
\ee
where $n$ is the electron density in a single stream. To compute 
$\varepsilon_L(\omega,{\bf k})$ we first perform integration by parts 
in Eq.~(\ref{eL}) and then, substituting the distribution function 
(\ref{f-2-streams}) into the resulting formula, we obtain in the 
nonrelativistic approximation
\be
\label{eL-3}
\varepsilon_L(\omega,{\bf k}) = 
\frac{\big(\omega^2 - ({\bf k} \cdot {\bf u})^2\big)^2
-2\mu^2 \big(\omega^2 + ({\bf k} \cdot {\bf u})^2\big)}
{\big(\omega^2 - ({\bf k} \cdot {\bf u})^2\big)^2} 
\;,
\ee
where ${\bf u}$ is the stream velocity (nonrelativistically 
${\bf u}= {\bf q}/m$ with $m$ being the electron mass) 
and $\mu^2 \equiv 4\pi e^2n/m$. There are four roots 
$\pm \omega_{\pm}({\bf k})$ of the dispersion equation 
$\varepsilon_L(\omega,{\bf k}) = 0$ which read 
\be
\label{roots}
\omega_{\pm}^2({\bf k}) = \mu^2 + ({\bf k} \cdot {\bf u})^2
\pm \mu \sqrt{\mu^2 + 4({\bf k} \cdot {\bf u})^2} \;.
\ee
As seen, $0 < \omega_+({\bf k}) \in R$ for any ${\bf k}$ but 
$\omega_-({\bf k})$ is imaginary for 
$({\bf k} \cdot {\bf u})^2 < 2 \mu^2$ when it represents the 
well-known two-stream electrostatic instability. For 
$({\bf k} \cdot {\bf u})^2 > 2 \mu^2$, the mode is stable,
$0 < \omega_-({\bf k}) \in R$. 

The correlation function 
$\langle E^i(\omega_1,{\bf k}_1) E^i(\omega_2,{\bf k}_2) \rangle$
as given by Eq.~(\ref{E-fluc-iso-long-2}) equals
\ba
\nonumber
\langle E^i(\omega_1,{\bf k}_1) 
E^i(\omega_2,{\bf k}_2) \rangle 
&=& - 16\pi^2 e^2 n \frac{(2\pi)^3\delta^{(3)}({\bf k}_1 + {\bf k}_2)}
{{\bf k}_1^2}
\Big[(\omega_1 + {\bf k}_1 \cdot {\bf u})
(\omega_2 + {\bf k}_2 \cdot {\bf u})
+ (\omega_1 - {\bf k}_1 \cdot {\bf u})
(\omega_2 - {\bf k}_2 \cdot {\bf u})\Big]
\\ [2mm] \nonumber 
&\times&
\frac{\omega_1^2 - ({\bf k}_1 \cdot {\bf u})^2} 
{\big(\omega_1 - \omega_-({\bf k}_1)\big)
\big(\omega_1 + \omega_-({\bf k}_1)\big)
\big(\omega_1 - \omega_+({\bf k}_1)\big)
\big(\omega_1 + \omega_+({\bf k}_1)\big) }
\\ [2mm] 
\label{E-fluc-2-stream-2} 
&\times&
\frac{\omega_2^2 - ({\bf k}_2 \cdot {\bf u})^2} 
{\big(\omega_2 - \omega_-({\bf k}_2)\big)
\big(\omega_2 + \omega_-({\bf k}_2)\big)
\big(\omega_2 - \omega_+({\bf k}_2)\big)
\big(\omega_2 + \omega_+({\bf k}_2)\big) }
\;.
\ea
One observes that the poles of the correlation function
$\langle E^i(\omega_1,{\bf k}_1) E^i(\omega_2,{\bf k}_2) \rangle$ 
at $\omega_1 = {\bf k}_1 {\bf v}$ and 
$\omega_2 = {\bf k}_2 {\bf v}$, which give the stationary 
contribution to the equilibrium fluctuation spectrum, have 
disappeared in Eq.~(\ref{E-fluc-2-stream-2}) as the
inverse dielectric functions vanish at these points.

The correlation function 
$\langle E^i(t_1,{\bf r}_1)E^i(t_2,{\bf r}_2) \rangle$ 
is given by Eq.~(\ref{EL-fluc-x}) with 
$\langle E^i(\omega_1,{\bf k}_1) 
E^i(\omega_2,{\bf k}_2) \rangle$ defined by 
Eq.~(\ref{E-fluc-2-stream-2}). Performing the trivial 
integration over ${\bf k}_2$ and taking into account
that $\omega_{\pm}(-{\bf k}) = \omega_{\pm}({\bf k})$,
one finds
\ba
\nonumber
\langle E^i(t_1,{\bf r}_1) E^i(t_2,{\bf r}_2) \rangle
&=& 32\pi^2 e^2 n \int_{-\infty +i\sigma}^{\infty +i\sigma}
{d\omega_1 \over 2\pi i}
\int_{-\infty +i\sigma}^{\infty +i\sigma}
{d\omega_2 \over 2\pi i}
\int {d^3k \over (2\pi)^3}
\frac{e^{-i \big(\omega_1 t_1 + \omega_2 t_2
- {\bf k}({\bf r}_1 - {\bf r}_2)\big)}}
{{\bf k}^2}
\big[\omega_1 \omega_2 - ({\bf k} \cdot {\bf u})^2 \big]
\\ [2mm] \nonumber 
&\times&
\frac{\omega_1^2 - ({\bf k} \cdot {\bf u})^2} 
{\big(\omega_1 - \omega_-({\bf k})\big)
\big(\omega_1 + \omega_-({\bf k})\big)
\big(\omega_1 - \omega_+({\bf k})\big)
\big(\omega_1 + \omega_+({\bf k})\big) }
\\ [2mm] 
\label{EL-fluc-x-stream1}
&\times&
\frac{\omega_2^2 - ({\bf k} \cdot {\bf u})^2} 
{\big(\omega_2 - \omega_-({\bf k})\big)
\big(\omega_2 + \omega_-({\bf k})\big)
\big(\omega_2 - \omega_+({\bf k})\big)
\big(\omega_2 + \omega_+({\bf k})\big)}
\;.
\ea
There are 16 contributions to the integrals over $\omega_1$ 
and $\omega_2$ in Eq.~(\ref{EL-fluc-x-stream1}) 
related to the poles at $\pm \omega_{\pm}$. Summing up the 
contributions, we get after lengthy calculation
\ba
\label{EL-fluc-x-stream6}
\langle E^i(t_1,{\bf r}_1) E^i(t_2,{\bf r}_2) \rangle
&=& 16\pi^2 e^2 n 
\int {d^3k \over (2\pi)^3} 
\frac{e^{i {\bf k}({\bf r}_1 - {\bf r}_2)}}{{\bf k}^2}
\frac{1}{(\omega_+^2 - \omega_-^2)^2}
\\[2mm] \nonumber
\times 
\bigg\{
\frac{\big(\omega_+^2 - ({\bf k} \cdot {\bf u})^2\big)^2}
{\omega_+^2}
&\Big[&
\big(\omega_+^2 - ({\bf k} \cdot {\bf u})^2\big)
\cos \big( \omega_+ (t_1 + t_2)\big)
+ 
\big(\omega_+^2 + ({\bf k} \cdot {\bf u})^2\big)
\cos \big(\omega_+ (t_1 - t_2)\big)
\Big]
\\[2mm] \nonumber
- 
\frac{
\big(\omega_+^2 - ({\bf k} \cdot {\bf u})^2\big)
\big(\omega_-^2 - ({\bf k} \cdot {\bf u})^2\big)} 
{\omega_+ \omega_- }
&\Big[&
\big(\omega_+ \omega_- - ({\bf k} \cdot {\bf u})^2\big)
\cos (\omega_+ t_1 + \omega_- t_2)
+
\big(\omega_+ \omega_- + ({\bf k} \cdot {\bf u})^2\big)
\cos (\omega_+ t_1 - \omega_- t_2)
\\[2mm] \nonumber
&+& 
\big(\omega_+ \omega_- - ({\bf k} \cdot {\bf u})^2\big)
\cos (\omega_- t_1 + \omega_+ t_2)
+
\big(\omega_+ \omega_- + ({\bf k} \cdot {\bf u})^2\big)
\cos (\omega_- t_1 - \omega_+ t_2)
\Big]
\\[2mm] \nonumber
+
\frac{
\big(\omega_-^2 - ({\bf k} \cdot {\bf u})^2\big)^2} 
{\omega_-^2}
&\Big[&
\big(\omega_-^2 - ({\bf k} \cdot {\bf u})^2\big)
\cos \big(\omega_- (t_1 + t_2)\big)
+
\big(\omega_-^2 + ({\bf k} \cdot {\bf u})^2\big)
\cos \big(\omega_- (t_1 - t_2)\big)
\Big]
\bigg\}\;.
\ea

Let us now consider the domain of wave vectors
$({\bf k} \cdot {\bf u})^2 < 2 \mu^2$ when 
$\omega_-({\bf k})$ is imaginary and it represents 
the unstable electrostatic mode. We write down 
$\omega_-({\bf k})$ as $i \gamma_{\bf k}$ with 
$0 < \gamma_{\bf k} \in R$,
\be 
\gamma_{\bf k} \equiv 
\sqrt{\mu \sqrt{\mu^2 + 4({\bf k} \cdot {\bf u})^2}
- \mu^2 - ({\bf k} \cdot {\bf u})^2} \;. 
\ee
We are interested in the contributions to the correlation function 
coming from the unstable modes. The contributions, which are 
the fastest growing functions of $(t_1+t_2)$ and $(t_1-t_2)$, 
correspond to the last term in Eq.~(\ref{EL-fluc-x-stream6}). 
The contributions provide
\ba
\label{EL-fluc-x-stream7}
\langle E^i(t_1,{\bf r}_1) E^i(t_2,{\bf r}_2) \rangle_{\rm unstable}
&=& 4\pi^2 e^2 n \int {d^3k \over (2\pi)^3} 
\frac{e^{i {\bf k}({\bf r}_1 - {\bf r}_2)}}
{{\bf k}^2
\mu^2 \big(\mu^2 + 4({\bf k} \cdot {\bf u})^2 \big)}
\frac{
\big(\gamma_{\bf k}^2 + ({\bf k} \cdot {\bf u})^2\big)^2} 
{\gamma_{\bf k}^2}
\\[2mm] \nonumber
&\times&
\Big[
\big(\gamma_{\bf k}^2 + ({\bf k} \cdot {\bf u})^2\big)
\cosh \big(\gamma_{\bf k} (t_1 + t_2)\big)
+
\big(\gamma_{\bf k}^2 - ({\bf k} \cdot {\bf u})^2\big)
\cosh \big(\gamma_{\bf k} (t_1 - t_2)\big)
\Big] \;,
\ea
where we have taken into account that
$\omega_+^2 - \omega_-^2 = 2\mu \sqrt{\mu^2 + 4({\bf k} \cdot {\bf u})^2}$.
As seen, the correlation function (\ref{EL-fluc-x-stream7}) is
invariant with respect to space translations -- it depends on the 
difference $({\bf r}_1 - {\bf r}_2)$ only. The initial plasma state
is on average homogeneous and it remains like this in course of
the system's temporal evolution. The time dependence of the correlation 
function (\ref{EL-fluc-x-stream7}) is very different from the space
dependence. The electric fields exponentially grow and so does 
the correlation function both in $(t_1 + t_2)$ and $(t_1 - t_2)$.
The fluctuation spectrum also evolves in time as the growth rate
of unstable modes is wave-vector dependent. After sufficiently
long times the fluctuation spectrum is dominated by the fastest 
growing modes. It should be remembered, however, that our
results hold for times which are not too long. Otherwise, the
perturbation, which exponentially grows, violates the condition
(\ref{smallness}) justifying the linearization procedure.

                                                                                
\section{Summary and conclusion}
\label{sec-discussion}
                                                                                

The calculations presented here show how to obtain a spectrum of 
electromagnetic fluctuations in equilibrium or nonequilibrium plasmas 
as a solution of an initial value problem. We first linearize the 
transport equation around the state which is on average stationary and 
homogenous. The linearized transport equation is solved together with 
the Maxwell equations by means of the one-sided Fourier transformation. 
The time dependent fluctuation spectrum is expressed through the  
fluctuations in the initial state. Electromagnetic initial fluctuations 
are determined by the initial fluctuations of the distribution function. 
The later are identified with the fluctuations in a classical system of
noninteracting particles. We compute fluctuation spectrum of longitudinal 
electric fields in isotropic plasma, and then there are considered 
fluctuations of magnetic and electric fields. Our equilibrium results 
coincide with those obtained by means of the fluctuation-dissipation 
theorem. However, the method adopted here clearly shows how the system 
looses its memory and how the stationary equilibrium spectrum of 
fluctuations emerges. As an example of unstable systems, the fluctuations 
of longitudinal electric field in the two-stream system are considered. 
The fluctuation spectrum appears to be qualitatively different than that 
of the equilibrium plasma.   

The scheme of calculation, which is worked our here in detail, can 
be applied to a variety of plasma nonequilibrium states. Our actual 
objective is, however, to generalize the scheme to study fluctuations 
in the quark-gluon plasma mentioned in the Introduction. Such a 
generalization is not quite trivial. The problem is that chromodynamic 
fields are, in contrast to their electromagnetic counterparts, gauge 
dependent, but physically meaningful correlation functions have to be 
gauge independent. A treatment of color charges needs to be different
as well. 


\end{document}